          [2001/04/25 1.1 (PWD)]
\documentclass[a4paper,twocolumn]{esapub} 
\usepackage{natbib}
\usepackage{graphicx}

\long\def\unmarkedfootnote#1{{\long\def\@makefntext##1{##1}\footnotetext{#1}}}

\title{Search for a light dark matter annihilation signal in the Sagittarius Dwarf Galaxy}
\author{B. Cordier$^{(1,0)}$}
\author{D. Atti\'e$^{(1)}$}
\author{M. Cass\'e$^{(1,3)}$}
\author{J. Paul$^{(1,4)}$}
\author{S. Schanne$^{(1)}$}
\author{P. Sizun$^{(1)}$}
\author{P. Jean$^{(2)}$}
\author{J.-P. Roques$^{(2)}$}
\author{G. Vedrenne$^{(2)}$}
\affil{$^{1}$ CEA-Saclay, DSM/DAPNIA/Service d'astrophysique, F-91191 Gif-sur-Yvette, France}
\affil{$^{2}$ Centre d'Etude Spatiale des Rayonnements, B.P. 4346, 31028 Toulouse Cedex 4, France}

\begin{document}


\maketitle

\begin{abstract}

The 511 keV emission from the Galactic Bulge observed by INTEGRAL/SPI could be the product of light (1-100 MeV) annihilating dark matter particles. In order to distinguish between annihilating light dark matter scenarios and more conventional astrophysical sources for the bulge emission, we propose to test the light dark matter hypothesis on the Sagittarius Dwarf Galaxy, a close-by galaxy dominated by dark matter. The predicted flux being in the SPI sensitivity range, the detection of a substantial 511 keV emission line from this galaxy 
would provide a strong evidence for the light dark matter hypothesis. 

During the two Galactic Center Deep Exposures performed in 2003, the Sagittarius Dwarf Galaxy was in the field of view of several INTEGRAL observations. In this paper we present preliminary results of the analysis of these data.

\end{abstract}
\section{Introduction}


\begin{figure*}
\centering
\includegraphics[width=0.8\linewidth]{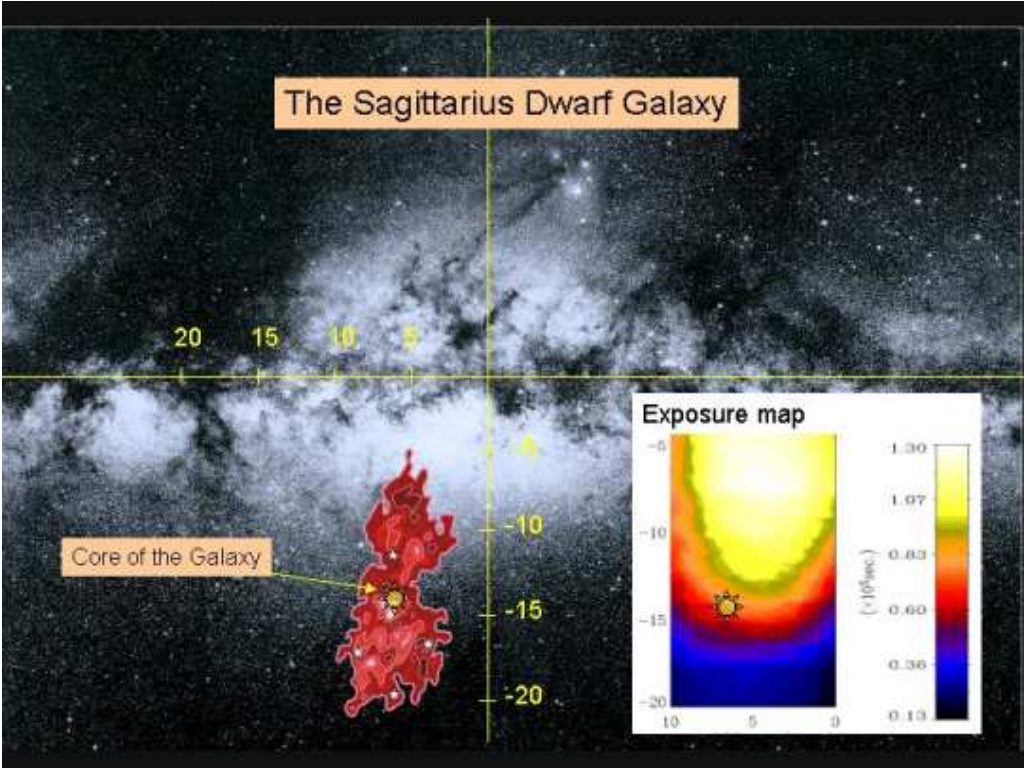}
\caption{Surface density map of the Sagittarius Dwarf Galaxy. On the right, Integral exposure map.\label{fig:expomap}}
\end{figure*}

The
\unmarkedfootnote{
corresponding author: b.cordier@cea.fr \\
$^{3}$ Institut d'Astrophysique de Paris, 98 bis Boulevard Arago, 75014 Paris, France \\
$^{4}$ F\'ed\'eration de Recherche Astroparticule et Cosmologie, 
Coll\`ege de France, 11 Place Marcellin Berthelot, 75231 Paris, France
} 511 keV emission from the Galactic bulge \citep{Jean03} observed by INTEGRAL/SPI \citep{Vedrenne03,Schanne02,Attie03,Roques03} could be the signature of light (1-100 MeV)
dark matter particles \citep{Casse04}. Such particles could annihilate throughout the Galactic bulge into positrons (and electrons)
which, after losing energy, annihilate into 511 keV gamma-rays \citep{Boehm04}.
Note that the SPI observation implies an injection rate of about $10^{43}$ positrons per second within the Galactic
bulge. It seems difficult to account for such an injection rate and to reproduce the morphology observed by
INTEGRAL/SPI by standard astrophysical scenarios such as Type Ia supernovae \citep{Schanne04}. 
However astrophysical scenarios cannot be ruled out and more evidence is needed to further motivate the light dark matter annihilation hypothesis.
In this context, \cite{Hooper04} have proposed to test the light dark matter hypothesis on the Sagittarius Dwarf
Galaxy (SDG), a closeby galaxy dominated by dark matter. A large dark matter annihilation rate is predicted from this
galaxy and as it contains comparatively few stars, the detection of a substantial 511 keV emission line from SDG
would provide a strong evidence for the light dark matter hypothesis.
 Indeed SGR is a good place to search for
the effect of pure dark matter since star formation has virtually stopped due to the lack of gas and thus supernovae and hypernovae should be essentially absent. 
Assuming that:

\begin{itemize}
\item all the 511 keV photons observed in the bulge of our Galaxy are the signature of dark matter annihilation
\item the spatial distribution of the SDG diffuse halo gas is less than 2 degrees
\item the shape of the dark matter halo profile for the SDG is similar to that of Draco, another dwarf galaxy
\item the hydrogen density in the SDG halo is of the order of ~ $10^{-4}~cm^{-3}$
\item the annihilation occurs directly on dust grains
\end{itemize}

\cite{Hooper04} found a predicted flux at 511 keV from SDG: (1-7)$\times10^{-4}~ph~cm^{-2}~s^{-1}$
. This work presents attempts to search for light dark matter annihilation in the SDG with SPI/INTEGRAL.

\section{The Sagittarius Dwarf Elliptical Galaxy}


The SDG is one of the most recently discovered members of the local group, and is currently in a very close encounter to our Galaxy. Its distance is approximately 25 kpc. It is apparently in the process of being disrupted by tidal gravitational forces induced by its big massive neighbor in this encounter.
Figure~\ref{fig:expomap} shows the position and the spatial distribution of the
SDG. The size is apparently big: 3$^{\circ}\times$8$^{\circ}$ in the sky. The core of the galaxy is located at $l$=5.6$^{\circ}$ and $b$= -14.0$^{\circ}$.
\cite{Ibata97} found that SDG orbits the Milky Way Galaxy in less than one billion years. Because it must have passed the dense central region of our Galaxy at least about ten times, it is surprising that the dwarf has not been completely disrupted so far. They suspect that this fact is an indication of significant amounts of dark matter within this small galaxy, which ties the stars more strongly to the galaxy by its gravity.

\section{The SPI/INTEGRAL observation }


The data analyzed in this work were accumulated during the two Galactic Centre Deep Exposures, executed as part of
INTEGRAL's guaranteed time observation. During these two programs, the SDG was in the instrument field of view at the
occasion of several successive pointings. Note that none of these pointings aimed exactly at the direction of the
SDG. However, taking into account the wide SPI fully coded field of view (16 degrees), we selected 121 of them with
an average exposure of 1400 seconds per pointing. Figure~\ref{fig:expomap} shows the INTEGRAL/SPI exposure map in the
region of the SDG for the two GCDEs. We note that the total effective observation time at the position of the SDG core is approximately 80~ks.

\section{Data analysis and results}


\begin{table*}
  \begin{center}
    \caption{Analysis results}\vspace{1em}
    \renewcommand{\arraystretch}{1.2}
    \begin{tabular}[h]{lll}
      \hline
      SPIROS 2$\sigma$ upper limit   & Model fitting 2$\sigma$ upper limit  & Predicted flux \\
      \hline
      4.8 $\times10^{-4}~ph~cm^{-2}~s^{-1}$ & 2.5 $\times10^{-4}~ph~cm^{-2}~s^{-1}$ & (1-7)$\times10^{-4}~ph~cm^{-2}~s^{-1}$ \\
      \hline 
      \end{tabular}
    \label{tab:table}
  \end{center}
\end{table*}

We assume that the spatial distribution of the source is less than 2 degrees. 
Thus, the source can be then considered as a point like source for SPI. 
The standard deconvolution method SPIROS is used for the extraction of spectral information.
The SPIROS method is described in \cite{Skinner03}. 
First, we run SPIROS in the imaging mode. 
This first step did not succeed to reveal any source. 
Then we built an input catalogue with only one source at the position of the SDG, 
and we run SPIROS in the spectral mode using the catalogue as an input. 
The combination of source intensity and background parameters, 
which best matches the data is found in each energy bin.

In order to cross check the SPIROS results, we analyzed again the data using the
model fitting method developed for the mapping of the Galactic Center. The data
preparation and analysis are described in \cite{Jean03} and \cite{Knodlseder03}. 
In this study, the spatial model used includes two components: 
a gaussian bulge component with a 10$^{\circ}$ FWHM, centered at $l$=0 and $b$=0, and a point source component located in $l$=6$^{\circ}$ and $b$=-14$^{\circ}$. 
The intensities of the components are fitted to maximize the likelihood provided by the spatial model and the background model to the observed distribution of counts for each 0.5 keV energy bin, each detector and each pointing. Along with the model intensities, 19 background model-scaling factors have been adjusted for the line component by the fit.
 The flux in the line is then derived by fitting a 3.7 keV FWHM gaussian to the spectrum resulting from the model fitting.

In both approaches, we do not detect the source. The results are presented in
table \ref{tab:table}. The upper limits are computed at a 2$\sigma$ confidence level, for a line width of ~3.7 keV
(the instrumental width of the Galactic center 511 keV line as measured by SPI).

\section{Discussion}


Even if SPI did not detect any 511 keV emission from the SDG, it should be remembered that the effective observation
time accumulated during the two GCDEs is only 80 ks. Thus, the flux upper limits derived on the SDG are still above the predicted fluxes. 

At this point it should be noted that the predicted fluxes are very optimistic. The calculation has been indeed performed assuming the most favourable hypotheses: 

1) \cite{Hooper04} consider that the spatial extension of the SDG halo is less
than 2 degrees and the shape of the halo profile for SDG is similar to that of
Draco. We know that the SDG is disrupted by tidal gravitational forces and that
the visible matter has an extent of 3$^{\circ}\times$8$^{\circ}$. 
In this context, the profile of the dark matter halo and the extent of the gas where the positrons annihilate can be
much larger than anticipated by the authors, leading to a size of the 511 keV source probably bigger than 2 degrees.
In such a context, we cannot further consider it as a point like source in our data analyses, the flux extraction will be more difficult, and the detection less favorable.

2) \cite{Hooper04} choose an hydrogen density in the SDG of the order of
$10^{-4}$~cm$^{-3}$. This parameter is critical for the computation of the thermalization time. The kinetic energy of the positrons is directly linked to the mass of the dark matter. For a dark matter mass greater than 10 MeV, the thermalization time could be larger than the age of the SGD.

3) In \cite{Hooper04}, the annihilation is supposed to occur directly on dust grains. If there is no dust,
the line will be broadened and the line width will be equal to 10 keV \citep{PVB03}. In this case again, the detection is less favourable.
Another interesting target would be Palomar-13 \citep{Casse04} provided this remote (24.3 kpc) globular cluster is not in a process of tidal disruption. The expected flux of 511 keV photons normalized on the galactic bulge emission using different dark matter distribution, is sufficiently high to be detected with INTEGRAL if Palomar-13 is really a dense clump of dark matter.

\section{Conclusions}


After two Galactic Center Deep Exposures, we have not detected a 511 keV emission from SDG, but this non-detection cannot exclude the light dark matter hypotheses. The SPI/INTEGRAL effective observation time is today not sufficient to reach the adequate sensitivity. In the following months and years, by accumulating the data, the SPI sensitivity in the direction of the SDG will reach the predicted fluxes. 
In this study we considered the 511 keV source as a point like source (less than 2 degrees), in the future we will search for different source extent. 
Moreover two parameters are very important for the prediction of the 511 keV emission as a signature of the light dark matter: the hydrogen density within the SDG and the dust fraction. Future observations from FUSE
will provide a more accurate value of the gas density.
With more SPI data and more accurate parameters, the observation of the SDG in the 511 keV energy range could permit to put a relevant constrain on the mass of the light dark matter.


\end{document}